\documentclass[11pt,twoside]{article}
\usepackage{asp2014}

\aspSuppressVolSlug
\resetcounters

\begin{document}

\title{It's your software! Get it cited the way you want!}

\author{Alice~Allen}
\affil{Astrophysics Source Code Library, Houghton, MI, US; \email{aallen@ascl.net}}
\affil{University of Maryland College Park, College Park, MD, US}

\paperauthor{Alice~Allen}{aallen@ascl.net}{0000-0003-3477-2845}{}{Author1 Department}{City}{State/Province}{Postal Code}{Country}



  
\begin{abstract}
Are others using software you've written in their research and citing it as you want it to be cited? Software can be cited in different ways, some good, and some not good at all for tracking and counting citations in indexers such as ADS and Clarivate's Web of Science. Generally, these resources need to match citations to resources, such as journal articles or software records, they ingest. This presentation covered common reasons as to why a code might not be cited well (in a trackable/countable way), which citation methods are trackable, how to specify this information for your software, and where this information should be placed. It also covered standard software metadata files, how to create them, and how to use them. Creating a metadata file, such as a CITATION.cff or codemeta.json, and adding it to the root of your code repo is easy to do with the ASCL's metadata file creation overlay, and will help out anyone wanting to give you credit for your computational method, whether it's a huge carefully-written and tested package, or a short quick-and-dirty-but-oh-so-useful code. 
  
\end{abstract}

\section{Why isn't your software being cited?}

Though not the only reasons your software might not be cited, here are four common ones for lack of citations: 

\begin{itemize}
    \item You haven't listed a preferred citation method, so others don't know that you expect your code to be cited when used, or don't know how to cite it
    \item Your preferred citation information is hard to find; see previous point as to why this is bad
    \item Your preferred citation isn't a trackable citation method; indexers cannot match the reference with the citation artifact you listed
    \item Your software isn't being used in research by others
\end{itemize}

\begin{figure}
    \centering
    \includegraphics{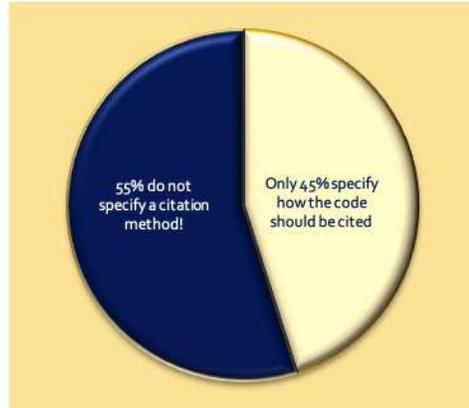}
    \caption{ASCL-registered codes with a preferred citation, 10/31/22}
    \label{fig:preferredcitationpercentage}
\end{figure}

Most software repos (\emph{e.g.}, GitHub, Bitbucket, GitLab) and web sites do not list a preferred citation method! As of October 31, 2022, only 45\% of the codes registered with the ASCL specify how that software should be cited, as Figure \ref{fig:preferredcitationpercentage} shows. This is despite the ASCL's repeated requests, made in email, on social media, and in presentations (such as this one!), that software authors clearly and explicitly list how their software should be cited, and make this information very easy to find. Most software sites do not include this important information!

\section{Software citation methods, good and bad}

Use of software in research can be acknowledged and cited in numerous ways, but they are not all equal. A footnote or mention in an acknowledgement or software section of an article is useful information for the reader of that article, but is not a formal citation and cannot be tracked by indexers such as ADS and Clarivate's Web of Science. Indexers track and count only resources that appear in a bibliography or reference section of a paper; in other words, there must be a formal citation, just as is done when citing a research article, for the software for that citation to be tracked \citep{2017ASPC..512..675A}. Not only must the citation appear in the bibliography, but it has to include certain correctly-formatted information for indexers to find it (again, just as is done for citing research articles), and the indexer has to be able to match that information with something it has already ingested \citep{henneken_edwin_2017_1011088}. This is why ADS, for example, indexes so many different journals: it uses information it has already ingested to track and match up information in newly-ingested articles. It cannot count what it is unable to match up, such as a citation to an article in a sports medicine journal. Table \ref{table:citationtrackability} lists different ways of acknowledging software use in astronomy research documents and whether indexers such as ADS can track and count this use as a citation; in the case of conference proceedings, this assumes the proceedings are formally published, and in all cases, this assumes that the bibliographic references are formatted correctly.

\begin{table}
\centering
\begin{tabular}{lc}
\hline
\textbf{Software citation method} & \textbf{Trackable?} \\ \hline
URL to the software in a footnote & NO!\\
Software listed in acknowledgements, no bibliographic reference & NO!\\
Code named in text, no bibliographic reference & NO!\\
Code named in text, bibliographic reference to proceedings paper & YES!\\
Code named in text, bibliographic reference to ASCL entry & YES!\\
Code named in text, bibliographic reference to Zenodo DOI & YES!\\
Code named in text, bibliographic reference to journal article & YES!\\ \hline
\end{tabular}
\caption{Which citation methods are trackable/countable?}
\label{table:citationtrackability}
\end{table}

\section{Increasing citations for your software}
If you want your research software used and cited by others, you can take concrete steps to help others find, understand, use, and properly acknowledge the contribution your computational method made to their research. These steps include:

\begin{itemize}
    \item Put your code online for download
    \item Assign your software a open license so others know how they can use your code
    \item Register your software with the Astrophysics Source Code Library (ascl.net) so others can find your software
    \item Provide a good README on your code site that explains why your software is useful
    \item Choose a trackable citation method and make your citation preference clearly visible and easy to find
    \item Cite your software when YOU use it
\end{itemize}
When deciding how you want your software cited, best practice per the FORCE11 Software Citation guidelines \citep{softwarecitationprinciples2016} is to cite the software directly, rather than an article. That said, these principles also recognize the value of citing a research paper that may provide additional information about the software and support citing an article \textit{in addition to} citing the software itself. Choose a trackable artifact as the citation method, such as the ASCL ID, Zeonodo DOI and, if you wish, a research article bibcode or DOI. 

Other steps you can take, too, that are more work (but very useful!) include providing a good tutorial on how someone new to your code can get started it using it, providing examples and documentation for your software, advertising it through events such as ADASS and other conferences, and creating a short video to demonstrate how your software enables research. 

\section{Two standard formats for citation information}
Software metadata --- information about the software itself, such as its title, URL for download, who authored it, and how it should be cited --- can be rendered in various ways. Two useful standard formats for representing the software metadata needed for citation are codemeta.json \citep{codemeta2017} and CITATION.cff \citep{citationcff2021}. The Codemeta standard uses a JSON file; this is machine-readable and is easy to reuse. These files can contain not only information needed to cite the software, but additional information useful to software archives, repositories, indexers, and journals. In contrast, CITATION.cff is contained in a YAML file; this is easy for humans to create and read, and contains only the information needed for citation. 

The ASCL can be used to generate codemeta.json metadata files, and as of this writing, it is expected that GitHub will also soon supply information and support for creating these files as well. To generate a codemeta.json file for an ASCL entry, add \textit{/codemeta.json} to an ASCL entry's URL (\textit{i.e.,} \url{https://ascl.net/1911.024/codemeta.json}).

GitHub offers instructions on creating and using CITATION.cff (see \url{https://tinyurl.com/y8r659h5}). If you use this method, please make sure your CITATION.cff does NOT specify the GitHub link as the citation method, as indexers cannot track and count GitHub URLs as citations. The ASCL can also create a CITATION.cff from an ASCL entry; the method is very similar to that listed in the paragraph above: add \textit{/CITATION.cff} to an ASCL entry's URL (\textit{i.e.,} \url{https://ascl.net/1911.024/CITATION.cff}). 

\begin{figure}
    \centering
    \includegraphics [scale=0.68]{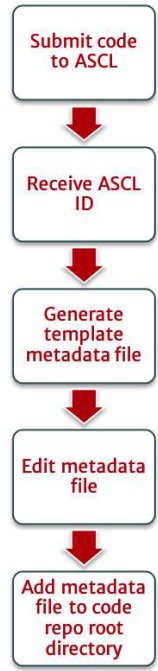}
    \caption{Creating a metadata file}
    \label{fig:metadatafileflowchart}
\end{figure}

Metadata files in these formats that are generated from an ASCL entry are intended to be a starting point; these files should be edited as needed by the software author to accurately represent the software and the preferred citation. Once a metadata file is created, it should be added to the root directory of your repo, as shown in Figure \ref{fig:metadatafileflowchart}, so others can easily find it.

\bibliography{P31}


\end{document}